\documentclass[11pt]{elsarticle}
\usepackage{geometry}
\usepackage{graphicx}

\usepackage{epsfig}
\usepackage{epstopdf}		

\usepackage{amsfonts,amsmath,amssymb,amscd}

\usepackage[figuresright]{rotating}

\usepackage{caption}
\usepackage{subcaption}

\usepackage[none]{hyphenat}

\usepackage{lineno}

\biboptions{round,sort&compress}

\newtheorem{theorem}{Theorem}[section]
\newtheorem{lemma}{Lemma}[section]
\newtheorem{proposition}{Proposition}[section]

\newtheorem{corollary}{Corollary}[section]
\newtheorem{example}{Example}[section]

\newcommand{\F}{\mathbb{F}}

\newcommand{\xn}{x^n - 1}

\begin{document}
	
\begin{frontmatter}

\title{On Quantum Codes Obtained From Cyclic Codes Over  $\F_2+u\F_2+u^2\F_2$ }

\author{Sukhamoy Pattanayak, Abhay Kumar Singh$^*$ and Pratyush Kumar}

\address{Department of Applied Mathematics, Indian School of Mines Dhanbad, India\\
	Email: sukhamoy88@gmail.com\\
	$^*$singh.ak.am@ismdhanbad.ac.in}

\begin{abstract}
Let $R=\F_2+u\F_2+u^2\F_2$ be a non-chain finite commutative ring, where $u^3=u$. In this paper, we mainly study the construction of quantum codes from cyclic codes over $R$. We obtained self-orthogonal codes over $\F_2$ as gray images of linear and cyclic codes over $R$. The parameters of quantum codes which are obtained from cyclic code over $R$ are discussed.
\end{abstract}

\begin{keyword}
 Quantum codes; cyclic codes; self-orthogonal codes; gray map. 
\end{keyword}

\end{frontmatter}
Mathematics Subject Classification  94B05. 94B15

\section{Introduction}
Quantum computation is the study of information processing tasks that can be accomplished using quantum
mechanical systems. This quantum security model is planned to explore error correction in quantum based systems. Quantum error-correcting codes have a prominent place in quantum communication as well as quantum computation. One of the most cogent arguments against the feasibility of quantum computation appears to be the difficulty of eliminating error caused by inaccuracy, decoherence and other quantum noise. Many good quantum cyclic error-correcting codes were constructed from Hamming codes, BCH codes and Reed-Solomon codes. The first quantum error-correcting code was discovered by Shor\cite{Sho}. \\
Cyclic codes over rings have generated a great deal of interest because of their rich algebraic structures and construction of quantum codes. Even through the theory of quantum error correcting codes has striking differences from the theory of classical error correcting codes, Calderbank et al.\cite{Cal} transformed the problem of finding quantum error correcting codes from classical error correcting codes over $GF(4)$ in . Many good quantum error-correcting codes have been constructed
by using classical cyclic codes over finite field $\F_q$ with self-orthogonal or dual containing properties (see Refs.\cite{Shi}). \\
Recently, the theory of quantum error-correcting codes over some special rings have
been developed rapidly. Several authors\cite{Qia,Qia1,Ash,Der} constructed quantum codes by using linear codes over finite rings . Qian et al.\cite{Qia1} gave a construction for quantum error correcting codes from cyclic codes of odd length over finite chain ring $\F_2 + u\F_2$ with $u^2=0$. Inspired by this study Kai and Zhu\cite{Kai} established a construction for quantum codes from cyclic codes of odd length over finite chain ring $\F_4 + u\F_4$ with $u^2=0$. They discussed Hermitian self-orthogonal codes over $F_4$ as Gray images of linear and cyclic codes over $\F_4 + u\F_4$. Existence condition of quantum codes which are derived from cyclic codes over finite ring $F_2 + uF_2 + u^2F_2, ~ u^3 = 0$ with Lee metric are discussed by Yin and Ma\cite{Yin}. Further, Qian\cite{Qia} gave a new method of constructing quantum error correcting codes from cyclic codes over finite ring $F_2 + vF_2,~ v^2 = v$, for arbitrary length. Later Ashraf and Mohammad\cite{Ash} provided the study quantum code from cyclic codes over $F_3+vF_3$, where $v^2=1$. In the recent paper\cite{Der}, good quantum codes obtained from cyclic codes over $A_2=F_2+uF_2+vF_2+uvF_2$. Motivated by Ashraf\cite{Ash} and Qian\cite{Qia}, we considered Cyclic Codes Over  $\F_2+u\F_2+u^2\F_2$, where $u^3=u$ to obtain quantum code. Cyclic code over finite chain ring have been study in several paper\cite{Abu,Sin}. Also cyclic code and self dual codes over $F_2 + uF_2$ have been extensively studied\cite{Bon}. Shi et al.\cite{Shi} discussed cyclic codes and weight enumerator linear codes over $\F_2+v\F_2+v^2\F_2$, where $v^3=v$.\\
The ring $R=\F_2+u\F_2+u^2\F_2,~u^3=u$ is a commutative non-chain ring.
The sequence of paper is structured as follows: In second section, we give some basic background about finite ring $R$, cyclic code and dual code. We define a Gray map from $R$ to $F_2^3$, and to show that if $C$ is self orthogonal so is $\Phi (C)$. We have also obtained the Gray image of cyclic code over $R$ in section 3. In section 4, we give a necessary and sufficient condition for cyclic codes over $R$ that contains its dual. The parameters of quantum error correcting codes are obtained from cyclic codes over $R$.

\section{Preliminaries}
Let $R$ be the commutative, characteristic 2 ring $R=\F_2+u\F_2+u^2\F_2 =\{a+ub+u^2c \vert a,b,c \in \F_2 \}$, with the property $u^3=u$. The principal ideal ring $R$ can also be thought of as the quotient ring $ \F_2[u]/\langle u^3-u \rangle$. It is endowed with obvious addition and multiplication. The unit elements of $R$ are $1$ and $1+ u+ u^2$.
Ideals of $R$ are listed below: \\
$\langle0\rangle $ $=$ $\left\lbrace 0 \right\rbrace,$\\
$\langle u \rangle$ $=$ $\left\lbrace 0, u, u^2, u+u^2 \right\rbrace,$\\
$\langle1+u\rangle$ $=$ $\left\lbrace 0, 1+u, u+u^2, 1+u^2 \right\rbrace,$\\   
$\langle u+u^2 \rangle$ $=$ $\left\lbrace 0,u+u^2 \right\rbrace,$\\
$\langle 1+u^2 \rangle$ $=$ $\left\lbrace 0,1+u^2 \right\rbrace.$\\
$R$ is a semi local ring with two maximal ideals $\langle u \rangle$ and $\langle 1+u \rangle$. A commutative ring $R$ is called a chain ring if its ideals form a chain under the relation of inclusion. From above, we can see that they do not form a chain as $\langle u \rangle$ and $\langle 1+u \rangle$ are not comparable. Therefore, $R$ is a non-chain ring.\\
A linear code $C$ of length $n$ over $R$ is a $R$-submodule of $R^n$. An element of $C$ is called a codeword.
A code of length $n$ is cyclic if the code is invariant under the automorphism $\sigma$ which has
\begin{center}
	$\sigma(c_0, c_1, \cdots , c_{n-1}) = (c_{n-1}, c_0, \cdots , c_{n-2}).$
\end{center}
It is well known that a cyclic code of length $n$ over $R$ can be identified with an ideal in the quotient ring $R[x]/\langle x^n-1\rangle$ via the $R$-module isomorphism
as follows:
\begin{center}
	$R^n \longrightarrow R[x]/\langle x^n-1\rangle$\\
	$(c_0,c_1,\cdots,c_{n-1}) \mapsto c_0+c_1x+\cdots+c_{n-1}x^{n-1} (\text{mod}\langle x^n-1\rangle)$
\end{center} 
The Hamming weight of a codeword $c$ is defined as $wt(c)$ $=$ $ \sum_{i=0} ^{n-1} wt(c_i)$, where 
\begin{equation}
wt(c_i) =\begin{cases}
1, & \text{ if $c_i \neq 0$}.\\
0, & \text{ if $c_i = 0$}.
\end{cases}
\end{equation}
The Hamming distance between two codewords $c^\prime$ and $c^{\prime\prime}$, where $c^\prime \neq c^{\prime \prime}$, is given by $d_H (c^\prime , c^{\prime \prime})$ $=$ $wt (c^\prime - c^{\prime\prime})$. Hamming distance of $C$ is defined as $d_H (C)$ $=$ $min$ $d_H ~(c^\prime , c^{\prime\prime})$, where $c^\prime,c^{\prime \prime}\in C$.
For any $x ~=~ (x_0,x_1,\ldots ,x_{n-1}),~ y ~=~ (y_0,y_1,\ldots,y_{n-1})$, the inner product is defined as
\begin{center} 
	$x\cdot y ~=~ \sum_{i=0}^{n-1} {x_i}{y_i}$.
\end{center}
If $x\cdot y=0$, then $x$ and $y$ are said to be orthogonal. The dual of a linear code $C$ over $R$ of length $n$ is given by 
$C^\perp = \left\lbrace x : ~ \forall ~ y \in C, x \cdot y = 0 \right\rbrace $.
It is also a linear code over $R $ of length $n$. A code $C$ is self-orthogonal if $C ~\subseteq~ C^\perp $, and self-dual if $C~=~C^\perp$.

A matrix $G$ is called the generator matrix of $C$, if its row vectors generates $C$ and all the row vectors are linearly independent. A linear code $C$ over R of length $n$ is also described in terms of its parity check matrix. A parity check matrix $H$ for the code $C$ is defined as\\
$C = \left\lbrace x \in C : Hx^T = 0 \right\rbrace $.
Two codes $C_1$ and $C_2$ having same length are said to be equivalent, if one can be obtained from the other by permuting the coordinates.
\begin{proposition}
	Any $R$-linear code containing some nonzero codewords is permutation equivalent to an $R$-linear code with a generator matrix $G$ of the form
	\[
	G= 
	\begin{pmatrix} 
	I_{k_1} & A & B_1 + u B_2 & C_1 + u C_2 & D_1 + u D_2 & E_1 \\
	0 & u I_{k_2} & 0 & (1+u) C_3 & u D_3 + (1+u) D_4 & E_2 \\
	0 & 0 & (1+ u) I_{k_3} & (1+ u) C_4 & (1+ u) D_5 & E_3 \\
	0 & 0 & 0 & (1+ u^2) I_{k_4} & 0 & E_4 \\
	0 & 0 & 0 & 0 & (u + u^2) I_{k_5} & E_5
	\end{pmatrix},
	\]
\end{proposition}
where ~~ 
$E_1 = E_{11} + u E_{12} + u^2 E_{13},\\
~ E_2  = u E_{21} + (1+ u) E_{22} + u^2 E_{23} + (1+ u^2) E_{24} + (1+u^2) E_{25},\\
~ E_3 = (1+ u) E_{31} + (1+ u^2) E_{32} + (u + u^2) E_{33}, \\
~ E_4 = (1+ u^2) E_{41} + (u + u^2) E_{42},\\
~ E_5 = (u+ u^2) E_{51};\\
~ A,~ B_1,~ B_2, ~ C_i~ {(i = 1 ~ \text{to}~ 4)} ,~ D_j ~{(j = 1 ~ \text{to}~ 5)}, ~E_{11}, ~E_{12}, ~E_{13}, ~E_{21}, ~E_{22}, ~E_{23},\\ ~E_{24}, ~E_{25}, ~E_{31}, ~E_{32}, 
~E_{33}, ~E_{41}, ~E_{42}, ~E_{51} $ are 2-ary matrices. And $C$ is an abelian group of type $8^{k_1}4^{k_2}4^{k_3}2^{k_4}2^{k_5}$  containing $2^{3{k_1}+2{k_2}+2{k_3}+{k_4}+{k_5}}$ codewords.\\
For a code $C$ to be self dual (i.e. $C=C^{\perp}$) over $R$, $k_2 = k_4$, $k_3 = k_5$, and $k_1 + k_2 + k_3 = n/2$, see reference\cite{Shi}. 

Let $g(x)$ and $h(x)$ be two polynomials in $R^n$ such that $g(x) h(x) $ $=$ $0$, we define the reciprocal polynomial of $h(x)$ to be $ \hat{h}(x) = x^{ deg(h(x))} h(x^{-1})$, its coefficients are those of $h(x)$ in reverse order. If $g(x)$ is a generator polynomial of $C$, then the generator polynomial
of $C$  is given by $\hat{h}(x)$, where $x^n - 1 ~=~ g(x) h(x)$.

\section{Gray Map}
Let $x=a+ub+u^2c$ be an element of $R$ where $a,b,c \in \F_2$. We define the Gray map $\Phi$ from $R$ to $\F_2 ^3$ is defined as $\Phi (a + u b + u^2 c) ~ = ~ (a, a + c, b) $ for all $ a, ~ b, ~ c ~ \in \F_2 $. It can be extended to map from $R^n$ to $\F_2 ^{3n}$ given by $\Phi : R^n 	\rightarrow \F_2 ^{3n}$ such that 
\[ \Phi (x_0, x_1, \ldots,x_{n-1})=(a_0,a_0+c_0,b_0, a_1,a_1+c_1,b_1,\ldots, a_{n-1},a_{n-1}+c_{n-1},b_{n-1})\] where $x_i=a_i+ub_i+u^2c_i$ for $i=0,1,\ldots.(n-1)$. 
From definition, the Lee weights of elements of $R$ are defined as follows\\
$w_{L}(0) = 0, ~ w_{L}(1) = 2, ~ w_{L}(u) = 1, ~ w_{L}( u^2) = 1, ~ w_{L}(1 + u) = 3, ~ w_{L}(1 + u^2) = 1, ~ w_{L}( u + u^2) = 2, ~ w_{L}(1+ u + u^2) = 2$.
Let $C$ be a linear code over $R$ of length $n$. For any codeword $c ~ = ~(c_0, c_1, \ldots,c_{n-1})$, the Lee weight of $c$ is defined as $w_L(c) = \sum^{n-1}_{i=0} {w_{L} (c_i)}$, where $w_{L} (c_i)$ denote the Lee weight of its $i th$ component. It is easy to verify that Lee weight of $C$ is the Hamming weight of its Gray image $\Phi (C) $.
Lee distance between two codewords $c$ and $c^\prime$, $ c \neq c^\prime $ is defined as $d_L (c,c^\prime) ~ = ~ w_L (c - c^ \prime) $.
Lee distance of $C$ is defined as $d_L(C) ~=  ~ min ~ d_L (c,c^\prime)$, $c$ and $c^\prime$ $\in C$.
\begin{theorem}
	The Gray map $\Phi$ is a distance preserving map from ($R^n$, Lee distance) to ($\F_2 ^{3n}$, Hamming distance). Moreover it is an isometry from $R^n$ to $\F_2 ^{3n}$.
\end{theorem}

\begin{theorem}
	If $C$ is a $(n,k,d_L)$ linear codes over $R$ then $\Phi (C)$ is a $(3 n, k,d_H)$ linear codes over $\F_2$, where $ d_H ~ = ~ d_L$.
\end{theorem}
\begin{pf}
	Let $x_1 ~ = ~ a_1 + u b_1 + u^2 c_2 $, $x_2 ~ = ~ a_2 + u b_2 + u^2 c_2 ~ \in ~ R, ~ \alpha ~ \in ~ F_2 $ then,\\
	$ \Phi(x_1 + x_2) = \Phi(a_1 + a_2 + u (b_1 + b_2) + u^2 (c_1 + c_2))$\\
	$~~~~~~~~~\quad \quad  = (a_1 + a_2, a_1 + a_2 + c_1 + c_2, b_1 + b_2)$\\
	$~~~~~~~~~\quad \quad  = (a_1, a_1 + c_1, b_1) + (a_2, a_2 + c_2, b_2)$\\
	$~~~~~~~~~\quad \quad  = \Phi (x_1) + \Phi (x_2)$\\
	$\Phi (\alpha x) = \Phi (\alpha a_1 + u \alpha b_1 + u^2 \alpha c_1)$\\
	$~~~\quad \quad  = (\alpha a_1, \alpha a_1 + \alpha c_1, \alpha b_1)$\\
	$~~~\quad \quad  = \alpha (a_1, a_1 + c_1, b_1)$\\
	$~~~\quad \quad  = \alpha \Phi (x)$\\
	so $\Phi$ is linear. Since $\Phi$ is bijective, therefore $  \mid C \mid = \mid \Phi (C) \mid $. From above theorem, we have $d_H = d_L$.
\end{pf}

\begin{theorem}
	Let $C$ be a code of length $n$ over $R$. If $C $ is orthogonal, so is $\Phi (C)$.
\end{theorem}
\begin{pf}
	Let $x_1 = a_1 + u b_1 + u^2 c_1, ~ x_2 = a_2 + u b_2 + u^2 c_2 $, where $a_1,~ b_1, ~c_1, ~ a_2, ~b_2, ~c_2 ~ \in ~\ F_2$.\\
	$x_1 \cdot x_2 = a_1 a_2 + u (a_1 b_2 + a_2 b_1 + b_1 c_2 + b_2  c_1) + u^2 (a_1 c_2 + a_2 c_1 + b_1 b_2 + c_1 c_2)$, \\
	if $C$ is orthogonal, then we have $a_1 a_2 = 0, ~ a_1 b_2 + a_2 b_1 + b_1 c_2 + b_2 c_1 ~=~ 0, ~ a_1 c_2 + a_2 c_1 + b_1 b_2 + c_1 c_2 ~=~ 0 $. Now,\\
	$\Phi (x_1) \cdot \Phi (x_2) ~ = ~ (a_1, a_1 + c_1, b_1) \cdot (a_2, a_2 + c_2, b_2)$\\
	$~~~~~~~~~ \quad \quad = ~ a_1 a_2 + a_1 a_2 + a_1 c_2 + a_2 c_1 + c_1 c_2 + b_1 b_2 $\\
	From above,
	$\Phi (x_1) \cdot \Phi (x_2) ~ = ~ 0$.\\
	Therefore, we have $\Phi (C) $ is self orthogonal.
\end{pf}

Let $c ~\in~\ F_2 ^{3n}$ with $c = (c_0, c_1, \ldots, c_{3n-1}) ~= ~ (c^{(0)} \mid c^{(1)} \mid c^{(2)}),$ where $ c^{(i)} ~ \in ~ \F_2 ^ {n}$ for $i = 0, ~1, ~2 $. And let $\lambda$ denote the cyclic shift from $\F_2 ^n$ to $F_2 ^n$ given by $\lambda (c^{(i)}) ~ = ~ ((c^{(i, n-1)}), (c^{(i, 0)}), \ldots, (c^{(i,n-2)}))$ for every $c^{(i)} ~=~  (c^{(i,0)}, c^{(i,1)}, \ldots, c^{(i,n-1)}) ~ \in ~ \F_2 ^n$, where $c^{(i, j)} ~\in ~ \F_2, ~ i ~=~ 0, ~1, ~2, ~and ~ j ~= 0, ~1,~2,~ \ldots, ~ n-1$. Let $\tau$ be a mapping from $\F_2 ^{3n} $ to $F_2 ^{3n}$  given by $\tau (c) = (\tau (c^{(0)}), \tau (c^{(1)}), \tau (c^{(2)}))$.

\begin{theorem}
	Let $\Phi$ be the gray map from $R^n$ to $\F_2 ^{3n}$, and $\lambda$ be the cyclic shift and $\tau$ be the mapping defined above. Then $\Phi \lambda ~ = ~ \tau \Phi$.
\end{theorem}
\begin{pf}
	Let $x_i ~=~ a_i + u b_i + u^2 c_i$ be the elements of $R$ for $i ~=~ 0,1,\ldots, n-1$. \\
	We have $\lambda (x_0, x_1, \ldots, x_{n-1}) ~ = ~ (x_{n-1}, x_0, x_1, \ldots, x_{n-2})$. \\
	Now we apply $\Phi$, we have \\
	$\Phi (\lambda (x_0, x_1, \ldots, x_{n-1})) ~ = ~ \Phi (x_{n-1}, x_0, x_1, \ldots, x_{n-2})$\\
	$~~~~~~~~~~~~~~~~~~\quad\quad\quad\quad = ~ (a_{n-1}, a_{0},\ldots, a_{n-2}, a_{n-1} + c_{n-1}, a_{0} + c_{0}, \ldots, a_{n-2} + c_{n-2}, b_{n-1}, b_{0}, \ldots, b_{n-2}) $. \\
	On the other hand \\
	$\Phi (x_0, x_1, \ldots, x_{n-1}) ~=~ (a_{0}, a_{1}, \ldots, a_{n-1}, a_{0} + c_{0}, a_{1} + c_{1}, \ldots, a_{n-1}+c_{n-1}, b_{0}, b_{1}, \ldots, b_{n-1})$\\
	By applying $\tau$, we have \\
	$\tau (\Phi  (x_0, x_1, \ldots, x_{n-1})) ~=~ (a_{n-1}, a_{0},\ldots, a_{n-2}, a_{n-1} + c_{n-1}, a_{0} + c_{0}, \ldots, a_{n-2} + c_{n-2}, b_{n-1}, b_{0}, \ldots, b_{n-2}) $.\\
	Thus, $\Phi \lambda ~ = ~ \tau \Phi$.
\end{pf}

\begin{theorem}
	A code $C$ of length $n$ over $R$ is cyclic iff $\Phi (C)$ is quasi cyclic code of index $3$ over $\F_2$ with length $3n$. 
\end{theorem}
\begin{pf}
	Suppose $C$ is cyclic code. Then $ \lambda (C) ~ = ~ C$. If we apply $\Phi$, we have $ \Phi (\lambda (C)) = \Phi (C)$. From above theorem, $\Phi (\lambda (C)) ~=~ \tau (\Phi (C)) = \Phi (C)$. Hence, $\Phi (C)$ is a quasi cyclic code of index $3$. Conversely, if $\Phi (C)$ is a quasi cyclic code of index $3$, then $\tau (\Phi (C)) = \Phi (C)$. From above theorem, we have $\tau (\Phi (C)) = \Phi (\lambda (C)) = \Phi (C)$. Since $\Phi $ is injective, it follows that $\lambda (C) ~=~C$.
\end{pf}

\section{Quantum code from cyclic codes over $R$}
Self-orthogonal code has an important application in the construction of quantum codes, as shown\cite{Cal} quantum error-correcting codes can be obtained from self orthogonal codes over $\F_2$.
\begin{theorem}
	Let $C$ and $C^\prime$ be binary $[n, k, d]$ and $[n, k_1, d_1]$ codes, respectively. If $C^{\perp} \subset C^\prime$, then an $[[n, k+k_1 -n, min \lbrace d,d_1 \rbrace]]$ code can be constructed. Especially, if $C^\perp \subseteq C$ then there exists an $[[n, 2k -n, d]]$ code.
\end{theorem} 

Let $C$ be a cyclic code of odd length $n$ over $R$. Since $R$ $=$ $\F_2+ u \F_2+u^2 \F_2 ~ =~ \F_2 [u] / \langle u (1 + u^2) \rangle$, for any element $ x = a + u b+ u^2 c \in R $, according to the Chinese Remainder Theorem, $ x = CRT^{-1}(a, a + b + c + b w)$, where $w = u + u^2$ and $w^2 = 0$.
Similarly,
\[  (a, A + B w) \xrightarrow{CRT} a + B u + (a + A + B) u^2 \in R \] 
where $A, ~B \in \F_2$. Hence, a code $C$ over $R$ can be written as $C~ =~ CRT^{-1}(C_2,C_w)$, where $C_2 \in \F_2, ~C_w \in R_w = \F_2 + w\F_2$. If $C_w$ is a cyclic code of odd length $n$ over $R_w$, then $C_w = \langle g(x), w a(x)\rangle = \langle g(x) + w a(x)\rangle$, where $g(x)$ and $a(x)$ are binary polynomials with $a(x) \vert g(x) \vert (x^n - 1) ~mod~2$.
\begin{theorem}\cite{Shi}
	Let $C$ be a cyclic code of odd length $n$ over $R$. Then $C$ is an ideal in $R_n=R[x]/\langle \xn \rangle$, which can be generated by $C ~=~ CRT^{-1}(C_2,C_w)$, where
	$C_2 ~=~\langle g_2 (x) \rangle,~ C_w ~=~ \langle g_1 (x) + w a_1 (x) \rangle,~ a_1(x) \vert g_1(x) \vert (x^n - 1),~ g_2(x)\vert (x^n - 1)$ and 
	$ \mid C \mid ~= ~  2^{3n - deg (g_1(x)) - deg (a_1(x)) - deg (g_2(x))}$.
	Moreover, there is a unique polynomial $g (x)$ such that $C= \langle g (x) \rangle$, namely, every ideal of $R_n$ is principal.
\end{theorem}
Since $g_1(x)\vert (x^n - 1)$, there exists $r_1(x) \in R_n$ such that $x^n - 1 =g_1(x) r_1(x)$, we set $g_w (x) = g_1(x) + w a_1(x)$, it follows that $x^n -1 = w g_w (x) r_1 (x)$, so $g_w(x) \vert (x^n - 1)$. Let $h_w (x) = (x^n - 1) /g_2 (x) = w  r_1(x)$. It is well known that, if $g(x)$ is generator polynomial of $C$, then the generator polynomial of $C^{\perp}$ is given by $ \hat{h}(x)$, where $\xn=g(x)h(x)$. We can obtain the following theorem   
\begin{theorem}\cite{Shi}
	Let $C$ be a cyclic code of odd length $n$ over $R$, which is generated by $C = CRT^{-1} (C_2,C_w)$, where \\
	$C_2 = \langle g_2 (x) \rangle, ~C_w = \langle g_1(x) + w a_1(x) \rangle, ~ a_1(x)\mid g_1(x) \mid (x^n -1), ~g_2 (x)\mid (x^n - 1)$.\\
	Then we have \\
	\[C^{\perp} = CRT^{-1} (C_2 ^{\perp} ,C_w ^{\perp}) ~=~ CRT^{-1} (\langle\hat{h}_2(x)\rangle, \langle w \hat{r}_1 (x)\rangle)\]
	and\\
	\[\mid C^{\perp} \mid ~=~ 2^{ deg (g_1(x)) + deg (a_1(x)) + deg (g_2(x))}\],\\
	where $h_2(x) = (x^n - 1)/ g_w (x)$.
	Moreover, there is a unique polynomial $ \hat{h}(x)$ such that $C^{\perp} = \langle\hat{h}(x)\rangle = \langle\hat{h}_2(x) + \hat{r}_1 (x) u + (\hat{h}_2 (x) + \hat{r}_1(x)) u^2 \rangle$.
\end{theorem}
The necessary and sufficient condition for self orthogonality of cyclic codes is given by lemma as follow:
\begin{lemma}\cite{Cal}
	A binary cyclic code $C$ with generator polynomial $g(x)$ contains its dual if and only if 
	\[ x^n - 1 ~\equiv ~ 0 ~(mod~ g(x) ~\hat{g}(x))  \],
	where $\hat{g}(x)$ is the reciprocal polynomial of $g(x)$.
\end{lemma}

\begin{lemma}\cite{Bon}
	Let $C = (f_1 f_2, w f_1 f_3)$ be a cyclic code of odd length $n$ over $R$, where $f_1 f_2 f_3 = x^n - 1$. Then $C$ is self-dual if and only if $f_1 = \hat{f}_3 $ and $ f_2 = \hat{f_2}$.
\end{lemma}
Now, we give a necessary and sufficient condition for cyclic code over $R$ that contains its dual.
\begin{theorem}
	Let $C = CRT^{-1} (C_2, C_w)$ be a cyclic code of odd length $n$ over $R$. Then $C^\perp \subseteq C$ if and only if 
	\[x^n - 1 \equiv ~0~ (mod ~g_2(x)~ \hat{g}_2(x))\]
	and 
	\[ x^n - 1 = f_1 f_2 f_3, \]
	where $\hat{g}_2(x)$ is the reciprocal polynomial of $g_2 (x)$ and $f_1 = \hat{f}_3 $ and $ f_2 = \hat{f_2}$, respectively.
\end{theorem}
\begin{pf}
	Let
	\[x^n - 1~ \equiv~ 0~ (mod ~g_2(x)~ \hat{g}_2(x))\]
	and 
	\[ x^n - 1 = f_1 f_2 f_3. \]
	Then by Lemma $1$ and $2$, we have $C_2 ^\perp \subseteq C_2$, $C_w ^\perp \subseteq C_w$, which implies
	\[CRT^{-1} (C_2 ^\perp, C_w ^ \perp) \subseteq CRT^{-1} (C_2, C_w)\]
	Therefore $C^\perp \subseteq C$.
	Next, if $C^\perp \subseteq C$, then  $CRT^{-1} (C_2 ^\perp , C_w ^\perp) \subseteq CRT^{-1} (C_2, C_w)$. From $CRT$, we get 
	\[ C_2 ^{\perp} \subseteq C_2, ~ C_w ^\perp \subseteq C_w \]
	Therefore
	\[ x^n - 1 \equiv 0 (mod ~g_2(x) \hat{g}_2(x))\]
	and 
	\[x^n - 1 = f_1 f_2 f_3. \]
\end{pf}

\begin{corollary}
	Let $C = CRT^{-1}(C_2, C_w)$ are cyclic code over $R$ of odd length $n$, then $C^\perp \subseteq C$, iff 
	\[C_2 ^\perp \subseteq C_2, ~C_w ^\perp \subseteq C_w. \]
\end{corollary}
From Theorem $6$ and $9$, we can now construct quantum code given by the following theorem.
\begin{theorem}
	Let $C = CRT^{-1} (C_2,C_w)$ be a cyclic code of odd length $n$ over $R$ with type $8^{k_1}4^{k_2}4^{k_3}2^{k_4}2^{k_5}$. If $C_2 ^\perp \subseteq C_2$ and $C_w ^\perp \subseteq C_w$, then $C^{\perp} \subseteq C$ and there exits a quantum error correcting code with parameters $[[3n, ~6 k_1+ 4(k_2 + k_3) + 2(k_4 + k_5) - 3n, ~ d_L]]$, where $d_L$ is the minimum Lee distance of $C$.
\end{theorem}

\begin{example}
	Let $n~=~3$, then 
	\[ x^3 - 1 ~=~ (x + 1)(x^2 + x + 1) \] 
	in $\F_2[x]$. Suppose $g_1(x)= g_2(x) = x + 1$ and $a_1(x)~=~ 1$, then $C = \langle g(x)\rangle = \langle 1 + u + u^2 +x \rangle$, $C$ is a linear cyclic code with length $3$ and Lee distance $2$. The dual code $C^\perp$ $=$ $\langle \hat{h}_2(x) + \hat{r}_1 (x) u + (\hat{h}_2 (x) + \hat{r}_1(x)) u^2 \rangle$ can be obtained of Theorem $8$. Clearly, $C^\perp \subseteq C$. Thus, we obtained a quantum code with parameters $[[9,5,2]]$. Suppose $g_1(x)= x+1, ~ g_2(x) = x^2 + x + 1$ and $a_1 (x) = 1$, then also in this case we find that $C^\perp \subseteq C$, and we obtained a quantum code with parameters $[[9,3,2]]$.
\end{example}

\begin{example}
	Let $n~=~5$, then 
	\[ x^5 - 1~=~ (x + 1)(x^4 + x^3 + x^2 + x + 1)\] 
	in $\F_2[x]$. Suppose $g_1(x) = g_2(x) = x + 1$ and $a_1(x)~=~ 1$, then $C= \langle g(x)\rangle ~=~ \langle 1 + v + v^2 + x \rangle$ is a linear cyclic code with length $5$ and Lee distance $2$. The dual code $C^\perp$ $=$ $\langle \hat{h}_2(x) + \hat{r}_1 (x) u + (\hat{h}_2 (x) + \hat{r}_1(x)) u^2 \rangle$ can be obtained of Theorem $8$. Clearly, $C^\perp \subseteq C$. Thus, we obtained a quantum code with parameters $[[15,11,2]]$.
\end{example}



\begin{thebibliography}{100}
	
	
	\bibitem{Abu} T. Abualrub and I. Siap, \emph{Designs,Codes and Cryptography} {\bf 42} (2007) 273-287.
	
	\bibitem{Ash} M. Ashraf and G. Mohammad,  \emph{Int.J. Quantum Inform.} {\bf 12} (2014) 1450042.
	
	\bibitem{Bon} A. Bonnecaze and P. Udaya, \emph{IEEE Trans. Inf. Theory} {\bf 45} (1999) 1250-1255.
	
	\bibitem{Cal} A. R. Calderbank, E. M. Rains, P. M. Shor and N. J. A. Sloane, \emph{IEEE Trans. Inf. Theory} {\bf 44} (1998) 1369-1387. 
	
	\bibitem{Der} A. Dertli, Y. Cengellenmis and S.Eren, \emph{Int.J. Quantum Inform.} {\bf 13} (2015) 1550031. 
	
	\bibitem{Kai} X. Kai and S. Zhu, \emph{Int.J. Quantum Inform.} {\bf 9} (2011) 689-700.
	
	\bibitem{Qia} J. Qian, \emph{Journal of Inform. and computational Science} {\bf 6} (2013) 1715-1722.
	
	\bibitem{Qia1} J. Qian, W. Ma and W. Gou, \emph{Int.J. Quantum Inform.} {\bf 7} (2009) 1277-1283.
	
	\bibitem{Shi} M. J. Shi, P. Sol´e and B. Wu, \emph{Appl. Comput. Math.} {\bf 12} (2013) 247-255.
	
	\bibitem{Sho} P. W. Shor, \emph{Phys. Rev. A} {\bf 52} (1995) 2493-2496.
	
	\bibitem{Sin}  A. K. Singh and P. K. Kewat, \emph{Designs, Codes and Cryptography} {\bf 74} (2015) 1-13.
	
	\bibitem{Ste}  A. M. Steane, \emph{Phys. Rev. A} {\bf 54} (1996) 4741-4751.
	
	\bibitem{Yin} X. Yin and W. Ma, \emph{International Joint Conferences of IEEE TrustCom-11} (2011).
	
\end{thebibliography}
\end{document}